# When AI Wears Many Hats: The Role of Generative Artificial Intelligence in Marketing Education

## ABSTRACT

Generative Artificial Intelligence (GAI) is increasingly being integrated into marketing education and is reshaping the skillsets required in marketing careers. While research has highlighted the promise and perils of incorporating GAI into education, there remains a need for a comprehensive framework to guide its effective use. In this research, we conduct a multipronged analysis, including a review of marketing course syllabi, a survey of marketing educators, and follow-up qualitative interviews. Building on Role Theory and the Community of Inquiry (CoI) model, we propose that GAI can assume three roles in marketing education: *tutor, teammate,* and *tool*. Each role influences *teaching, social,* and *cognitive* presence differently, shaping the learning experience and preparing workplace-ready marketing graduates. For instance, as a *tutor*, GAI can aid students in grasping theoretical concepts, while as a *teammate*, it can foster collaboration by supporting brainstorming and problem-solving activities. However, ethical considerations such as data privacy, plagiarism, dependency on AI, and fairness in assessment must be addressed to ensure its responsible adoption in marketing education. We provide concrete examples for GAI's careful integration in marketing courses, and its implications for marketing educators, learners, and policymakers.

## INTRODUCTION

Generative Artificial Intelligence (GAI) is reshaping marketing education and jobs, creating new opportunities and challenges (Grewal et al. 2024a; Grewal et al. 2024b; Kumar et al. 2024). GAI technologies like ChatGPT—a conversational AI model that generates human-like responses—are increasingly being integrated into marketing courses. These technologies have the potential to impact learner engagement across diverse marketing topics (e.g., consumer behavior, digital marketing) and skills (e.g., critical thinking, effective communication). On the one hand, integrating GAI into marketing curricula could provide students with a valuable learning experience and a competitive edge in the job market by familiarizing them with industry-relevant AI technologies (Dubey, Astvansh, and Kopalle 2024; Forsey 2023; Torres 2024). For example, GAI can assist students in learning marketing concepts like customer segmentation through personalized feedback or support the development of product prototypes by functioning as a creative collaborator. On the other hand, increased reliance on GAI could diminish students' ability to develop foundational skills independently, such as problem-solving and deep analytical thinking. The widespread adoption of GAI in education and the workforce may lead to concerns about job displacement, ethical implications, and the need for continual adaptation to these rapidly evolving technologies (Ding, Dong, and Grewal 2024).

Despite its opportunities and challenges, most marketing educators currently use GAI through a "trial-and-error approach" rather than established best practices (Guha, Grewal, and Atlas 2024). While prior research has highlighted GAI's potential to enhance student autonomy, creativity, and enjoyment (Rindfleisch, Kim, and Lee 2024), questions remain about how these technologies should be deployed to optimize learning outcomes. Emerging research suggests that GAI can provide unbiased and consistent feedback in tasks such as peer evaluations and case study analysis, yet its effectiveness may be limited by the lack of a cohesive framework for its use and related ethical issues (Jürgensmeier and Skiera 2024). Our research examines how marketing education can integrate GAI in various roles to support diverse learning objectives in marketing courses while ensuring students' preparedness for an AI-driven workforce.

Our research adopts a unique theoretical lens grounded in Role Theory (Biddle 1986) and the Community of Inquiry (CoI) framework (Garrison, Anderson, and Archer 1999) to conceptualize the roles GAI can play in marketing education. Specifically, we propose that GAI can act as a *tutor, teammate,* or *tool*, depending on the learning objectives of the marketing course. For example, as a *tutor*, GAI might provide guidance on lower-order learning tasks such as understanding key metrics like customer lifetime value. As a *teammate*, it can collaborate on higher-order creative tasks, such as brainstorming ideas, designing marketing campaigns or product prototypes. Finally, as a *tool*, GAI can facilitate data analysis and other operational tasks, freeing up students to focus on strategic tasks and critical thinking. By incorporating GAI in these roles into both core and elective courses, marketing educators may be able to achieve the learning objectives of their courses more effectively. In core courses, where foundational skills like analysis and critical thinking are emphasized, GAI can support lower-order learning objectives as a *tutor* or *tool*. In elective courses, which often prioritize creativity and specialized skills, GAI's role as a *teammate* may better align with higher-order learning objectives. When these roles are mapped onto the CoI model, GAI can influence educational outcomes through *teaching presence* (instructor-facilitated learning), *social presence* (collaborative learning), and *cognitive presence* (critical engagement with content). By integrating these perspectives, our research provides a comprehensive framework for understanding and optimizing GAI's evolving role in marketing education.

Our research offers a commentary on three key themes relevant to the integration of GAI in marketing education. First, we uncover the distinct roles that GAI can play in marketing education, shaped by the learning objectives and topics covered in a course. Second, we discuss how these roles influence educational experiences through the dimensions of *teaching, social,* and *cognitive* presence. Finally, we highlight the broader implications of GAI's use from the perspective of marketing students, considering their learning outcomes and workplace readiness.

To explore these themes, we take a multipronged approach that combines a review of the education, marketing, and AI literature, an analysis of the syllabi from diverse marketing courses to identify current practices, a survey of marketing faculty at a major U.S. public university, and qualitative discussions with marketing educators to capture practitioner perspectives (Alba and Hutchinson 1987). Building on our insights, we provide detailed examples for marketing educators to integrate GAI effectively and offer an agenda for future research. We emphasize the need for policy guidelines that address ethical considerations, such as privacy concerns and equitable access to GAI technologies for students.

Our findings have significant implications for both academics and policymakers. We provide actionable strategies for marketing educators to incorporate GAI responsibly as a *tutor, teammate,* and *tool,* ensuring that students acquire not only technical proficiency but also the ethical and critical thinking skills for navigating AI-driven marketing roles in the workplace. For policymakers, we highlight the need for guidelines that balance innovation with accountability, addressing concerns such as data privacy, algorithmic bias, and equitable access to GAI technologies. By offering a framework for integrating GAI into marketing education, we contribute to the broader discourse on preparing marketing graduates for an AI-driven future.

## CONCEPTUAL DEVELOPMENT: ROLE OF GAI IN MARKETING EDUCATION

While AI is a broad term encompassing the development of computer systems that are able to perform tasks typically requiring human intelligence, GAI refers to a branch of AI comprising systems that can create new content, including text, images, and other forms of data. In marketing education, GAI encompasses relevant technologies, such as large language models (e.g., ChatGPT) and other AI-powered technologies that can generate, manipulate, or analyze marketing-related content. These systems can assist in various educational tasks, from content creation to data analysis and personalized tutoring. Consistent with these definitions, we use the term "GAI" in this paper to refer to generative AI technologies and the term "AI" to refer to artificial intelligence more broadly.

To understand the role of GAI in marketing education, we integrate relevant theoretical perspectives in our research. Specifically, we draw from Role Theory and the Community of Inquiry (CoI) model (Biddle and Thomas 1966; Garrison, Anderson, and Archer 1999). Recent research on AI in marketing education has acknowledged the roles played by GAI, highlighting the importance of a role perspective informally (Guha, Grewal, and Atlas 2024; Mollick and Mollick 2023). Separately, the CoI model in the education literature has provided a framework for improving learning outcomes through educational experience based on *teaching, social,* and *cognitive* presence, which are achieved by the roles instructors and students play to create a community of inquiry. In this section, we integrate the two perspectives to understand how GAI can take on various roles in marketing courses and impact learners' educational experience. We draw on the research on AI in marketing education and the scholarship of teaching and learning (SoTL) to inform our conceptualization (see Table 1 for a review of marketing research).[1]

---

[1] In Web Appendix A, Table A1, we provide an additional review of relevant papers in other disciplines, such as economics and information systems.

**Table 1. Review of Relevant Literature**

| | Authors (Year) | Research Focus | Research Method | Research Gap |
|---|---|---|---|---|
| **SoTL Literature Review in Marketing Education** | Bacon and Stewart (2022) | Actual learning in marketing and management education | Systematic review | Current marketing education tools may not improve the efficacy of marketing education in the GAI era.<br><br>No integrative framework for GAI use by marketing educators that considers the roles played by AI, learners' educational experience, and work readiness in the GAI era. |
| | Crittenden (2023) | Scholarship of teaching and learning in marketing education | Conceptual | |
| | Crittenden (2024) | Future of marketing education; implications for teaching and learning | Conceptual | |
| | Crittenden, Biel, and Lovely (2019) | Embracing digitalization for student learning and new technologies | Conceptual | |
| | Freeman et al. (2017) | "Work-ready" marketing graduate | Conceptual | |
| | Goli, Chintagunta, and Sriram (2022) | Payment and user Engagement in online courses | Empirical, quantitative | |
| | Greenacre et al. (2017) | "Work-Ready" Marketing Graduate | Conceptual | |
| | Honea, Castro, and Peter (2017) | Marketing competencies and workplace readiness | Conceptual | |
| | Kim et al. (2022) | Home-Tutoring Services and AI technology | Experiment, quantitative | |
| | Ko et al. (2023) | AI-powered education app learning outcomes | Empirical, quantitative | |
| | Lee, Lee, and Jeong (2023) | Digital textbooks and student performance | Empirical, quantitative | |
| | McArthur et al. (2017) | Work-readiness and marketing job advertisements | Content analysis | |
| | Mitchell, Rippe, and Kemp (2022) | Teaching turmoil and triumphs in times of crisis. | Conceptual | |
| | Narang, Yadav, and Rindfleisch (2022) | Content sharing and engagement in online learning platforms | Experiment, quantitative | |
| | Rohm, Stefl, and Ward (2021) | Live project-based learning and students' skill development | Conceptual | |
| | Zhou et al. (2021) | Machine learning and courseware consumption | Machine Learning, quantitative | |
| ***Research on Gen AI and Marketing Education and Policy*** | Acar (2024) | AI-first strategy and firms' practices | Conceptual | Lack of a unified framework to guide marketing educators and policymakers on using GAI across roles (tutor, teammate, tool) and learning objectives (lower and higher order), for teaching, cognitive, and social presence in marketing education and work readiness. |
| | Bell et al. (2024) | AI and idea generation | Machine Learning, quantitative | |
| | Bluvstein et al. (2023) | Human-like error and information disclosure | Lab experiment | |
| | Ding, Dong, and Grewal (2024) | GAI and marketing classroom | Conceptual | |
| | Grewal, Guha, and Becker (2024) | The impact of GAI for policy and society | Conceptual | |
| | Guha, Grewal, and Atlas (2024) | GAI and marketing education | Conceptual, survey | |
| | Kumar et al. (2024) | GAI and policy implications of its use by businesses | Conceptual, qualitative | |
| | Madhavaram and Appan (2025) | Enabling AI, engaging AI, and ersatzing AI's impact on marketing | Conceptual | |
| | Narang, Ahlbom, and Banerjee (2024) | GAI and marketing education | Empirical, quantitative | |
| | Peres et al. (2023) | GAI and marketing field | Conceptual | |
| | Rindfleisch, Kim, and Lee (2024) | GAI and convivial tool | Conceptual, field study | |
| | This research (2025) | GAI as a teacher, teammate, and tool, and its impact on marketing education through teaching, cognitive, and social presence | Conceptual, survey, interviews | Guides theory and practice for using GAI in marketing education |

Traditionally applied separately in educational research, the integration of Role Theory and the CoI model can expand our understanding of GAI in marketing education in at least two ways. First, Role Theory has been applied to understand human behavior based on the context and social position but not in the case of roles assumed by AI, an important emerging actor in the learner-instructor dynamic (Latour 2007). Second, the CoI model, which is widely used for studying learning processes in online and blended environments, has been associated with deeper learning and higher-order thinking among students (Ariati, Pham, and Vogler 2023), serving a key goal of marketing education for creating job-ready graduates (Freeman et al. 2017; Greenacre et al. 2017). Despite their potential to expand our understanding of the role of GAI in marketing education, these two theoretical perspectives have not been integrated so far in the literature.[2]

GAI can take on roles as "an assistant that can improve writing abilities, a (smart) colleague that can ask probing questions or help with ideation, or an expert that can clarify technical concepts" in marketing courses (Guha, Grewal, and Atlas 2024). GAI in marketing courses can also be viewed as a convivial tool that can improve autonomy, empowerment, and enjoyment of learners (Rindfleisch, Kim, and Lee 2024). Alternatively, GAI can be classified on the basis of its function as enabling, engaging, and ersatzing (artificial, but inferior intelligence that makes marketing entities less capable, Madhavaram and Appan 2025). GAI can play many roles in the classroom and faculty can leverage it appropriately in these roles (Mollick and Mollick 2023). Despite its benefits, GAI also poses many challenges. It may perpetuate biases or inaccuracies (Acar 2024; Ding, Dong, and Grewal 2024) and can produce overly similar or uninspired ideas if overused (Bluvstein et al. 2023; Dietvorst, Simmons, and Massey 2018;

[2]We provide a detailed overview of Role Theory and the CoI model in Web Appendix B.

Narang, Ahlbom, and Banerjee 2024). Emerging studies discuss AI tools like ChatGPT in marketing research and education (Peres et al. 2023) and highlight GAI's role in creativity and idea generation (Bell et al. 2024).

Based on our review, we describe and illustrate with examples the three primary roles GAI can play when used effectively in marketing education. These appear in Table 2.

**Table 2. Roles Assumed by GAI in Marketing Education**

| GAI Role | Role Description | Example from Marketing Courses | Supporting Literature |
|---|---|---|---|
| Tutor | As a tutor, GAI provides personalized instruction and helps with learning specific concepts and skills. | Principles of Marketing: GAI offers personalized feedback on student projects, guiding them through marketing strategies and concepts. | Bal et al. (2015); Ding, Dong, and Grewal (2024); Guha, Grewal, and Atlas (2024) |
| Teammate | As a teammate, GAI collaborates by sharing ideas and offering different perspectives in discussions. | Marketing Research: GAI facilitates brainstorming sessions, providing alternative perspectives on research questions and methods. | Meincke et al. (2024); Hwang et al. (2020) |
| Tool | As a tool, GAI completes assigned tasks efficiently, following commands as directed. | Marketing Analytics: GAI helps apply data analytics techniques and generates code and reports. | Jürgensmeier and Skiera (2024); Mollick and Mollick (2023); Rindfleisch, Kim, and Lee (2024) |

As shown in Table 2, GAI can play the role of a marketing *tutor*, *teammate*, or *tool* in marketing courses. As a *tutor*, GAI offers personalized instruction, such as providing feedback on student projects in Principles of Marketing courses (Bal et al. 2015; Ding, Dong, and Grewal 2024). As a *teammate*, GAI contributes to collaborative tasks, such as aiding brainstorming sessions in Marketing Research (Meincke et al. 2024; Hwang et al. 2020). Lastly, GAI functions as a *tool* in courses like Marketing Analytics, performing tasks like data analysis and report generation (Jürgensmeier and Skiera 2024).

Role Theory enables an analysis of how GAI reshapes expectations for educators and students depending on the role for which the AI is employed. On the one hand, instructors could fully transition from primary content providers to facilitators of AI-assisted learning. On the other hand, students exposed to GAI technologies could shift from passive information recipients to active collaborators with GAI in their educational journey depending on the role played by the GAI. More generally, roles tend to be dictated by expectations and norms (Biddle 1986; Whetten 2021). In the case of marketing education, the expectations are drawn from the nature of the course, such as course level (e.g., 400-, 500-), type (e.g., core, elective), learning objectives (e.g., high-, low- order), topics (e.g., simple, complex), and student familiarity with AI. Therefore, the role of AI in marketing education could depend on the course and its goals (Bloom et al. 1956).

The CoI model defines a learner's educational experience as comprising three core elements—*teaching* presence, *social* presence, and *cognitive* presence (Garrison, Anderson, and Archer 1999). The three core dimensions each contribute to the educational experience and will be impacted depending on the use of GAI in different roles in a marketing course (Wang et al. 2024). GAI can be introduced in marketing courses in multiple roles depending on the learning objectives and marketing topics. The SoTL literature emphasizes that effective course design requires a strategic blend of pedagogical theory and practical application, particularly in applied fields like marketing (Arbaugh, Bangert, Cleveland-Innes 2010) and that courses with clear outcomes and active learning strategies enhance critical thinking and application of marketing concepts and skills (Bacon and Stewart 2022; Bacon et al. 2008; Crittenden 2023). Our research relates to the research on marketing course design and delivery in SoTL by focusing on the learning objectives of marketing courses (e.g., to understand concepts, to apply tools) as a starting point for integrating GAI into course design and delivery in various roles.

In Table 3, we posit ways in which GAI in its various roles may influence *teaching, social,* and *cognitive* presence with examples from marketing courses.

When GAI acts as a *tutor*, its primary impact is on *teaching* presence, offering personalized guidance and immediate feedback, such as explaining marketing concepts (e.g., customer segmentation) or creating review materials (e.g., glossary flashcards with key

**Table 3. How GAI Roles Might Impact Teaching, Social, and Cognitive Presence: Illustrative Examples from Marketing Courses**

| GAI Role/ CoI | Teaching Presence | Social Presence | Cognitive Presence |
|---|---|---|---|
| Tutor | Impacts *teaching* presence by guiding students through concepts (e.g., explaining customer segmentation or brand positioning in a Principles of Marketing course). | Limited or no impact on *social* presence, as the interaction is typically one-on-one with the GAI serving as an expert tutor. | Impacts *cognitive* presence by explaining the content and key concepts (e.g., creating a review guide or glossary of terms in Consumer Behavior or Market Research courses). |
| Teammate | Impacts *teaching* presence through monitoring and giving feedback on group projects (e.g., brainstorming new project ideas in a Marketing Analytics course). | Impacts *social* presence by engaging with students in group-based tasks (e.g. helping a team identify target market segments for a Marketing Management course simulation). | Impacts *cognitive* presence by encouraging critical thinking and creative ideation during brainstorming sessions (e.g., assisting students in refining their research hypotheses, like creating a pricing strategy for a new product). |
| Tool | Impacts *teaching* presence by automating tasks or demonstrating a technique learners need (e.g., analyzing marketing data, generating reports from survey data in a Marketing Research course) | Limited or no impact on *social* presence, as the interaction is typically one-on-one with the GAI serving as a tool to execute tasks. | Impacts *cognitive* presence indirectly by freeing up mental resources to focus on strategic and higher-order rather than automatable tasks (e.g., analyzing data so students can focus on interpreting and applying the results in a Marketing Research course). |

definitions) in a Principles of Marketing course (Jürgensmeier and Skiera 2024). However, over-reliance on GAI can weaken *teaching* presence by reducing human interaction, potentially limiting mentorship opportunities. Its impact on *social* presence is minimal, as interactions are typically one-on-one with the AI, but *cognitive* presence may be enhanced through deeper understanding and reinforcement of key ideas. For instance, platforms like Khanmigo by Khan

Academy can deliver tailored, interactive learning experiences by adapting to individual students' needs. Another example of GAI's tutoring capabilities is its ability to automate the creation of lecture videos in marketing courses. By training AI on an instructor's style, voice, and persona using pre-recorded clips, and feeding it a script, the AI can generate complete lecture videos within seconds, closely mimicking the instructor's delivery (Byrne 2023).

When GAI acts as a *teammate*, it primarily influences *social* presence, facilitating collaboration during group projects, such as brainstorming target market segments for a Marketing Management course simulation. It can also foster *cognitive* presence by encouraging students to think critically while generating and refining ideas, like developing a pricing strategy for a new product. However, over-reliance on GAI risks diminishing peer-to-peer interaction and creating dependencies that weaken collaborative skills. One example of GAI's collaborative capabilities as a teammate is Yellowdig AI, which facilitates online learning communities by encouraging discussion, feedback, and idea-sharing among students while integrating AI to enhance participation and engagement through gamification (Akchurina and Albuquerque 2024; Yellowdig 2024). In marketing courses, Yellowdig with AI capabilities can improve learners' ease of posting content (Narang, Ahlbom, and Banerjee 2024).

When GAI functions as a *tool*, it predominantly impacts *cognitive* presence by automating repetitive tasks and enabling students to focus on higher-order analysis, such as interpreting data results in a Marketing Research course. The role also supports *teaching* presence, as instructors use GAI to demonstrate techniques or automate feedback. While its influence on *social* presence may be limited, it could indirectly improve group interactions by freeing up cognitive resources. For instance, a course-specific chatbot trained on syllabi can address individual student queries instantly, keeping them engaged with the material while freeing up time and resources for more involved group tasks with peers (NCSA 2023).

While our framework delineates distinct roles and their primary influences on *teaching, social,* and *cognitive* presence, it's important to acknowledge that in practice, these roles often overlap and interact dynamically, particularly in online and hybrid formats (Crittenden, Biel, and Lovely 2019). For instance, in a blended marketing analytics course, GAI might simultaneously function as a tool for data analysis and a teammate for collaborative problem-solving, while also providing tutoring support through personalized feedback. This multi-faceted deployment can create synergistic effects across different types of presence. When GAI assists with routine calculations (tool role), it not only enhances *cognitive* presence through task automation but may also strengthen *teaching* presence by allowing instructors to focus on higher-order concept explanation. Similarly, in group projects, GAI might transition fluidly between teammate and tool roles, facilitating both social presence through collaborative ideation and *cognitive* presence through technical support (e.g., Lee, Lee, and Jeong 2023; Narang, Yadav, and Rindfleisch 2022). Research on AI-powered learning platforms significant improvement in online learner engagement across multiple metrics (Ko et al. 2023; Zhou et al. 2021). However, Kim et al. (2022) noted variability in tutor adoption of AI, with minimal performance benefits. Early research on GAI in education remains limited, offering little theoretical guidance for its broader use in marketing education (Narang, Ahlbom, and Banerjee 2024). Notably, the potential of GAI to enhance asynchronous learning by providing real-time, adaptive feedback could help bridge the gap between synchronous and asynchronous experiences, making online courses feel more interactive and engaging.

Integrating the CoI model with Role Theory provides a comprehensive lens to analyze these complex interactions in marketing education. While GAI's roles influence each type of presence to a different degree, there may also be potential spillover effects (Kozan and Richardson 2014). For example, effective *teaching* presence can bolster *cognitive* presence through enhanced instruction quality, while strengthened *social* presence fosters engagement that enriches the overall educational experience. These interactions become particularly salient in hybrid teaching formats, where GAI must adapt to both online and in-person learning contexts. Our framework helps educators strategically align GAI's multiple roles with course objectives while maintaining active human involvement and minimizing potential downsides (Arbaugh and Hwang 2006; Arbaugh 2008). Understanding these role overlaps and their impacts across different types of presence is crucial for marketing educators designing integrated learning experiences that leverage GAI's full potential across different instructional modalities.

Importantly, our framework can guide the use of GAI in marketing courses for specific needs and skills to prepare marketing graduates to meet workplace demands (Freeman et al. 2017; Greenacre et al. 2017; McArthur et al. 2017). Marketing practitioners evaluate job applicants' competencies using a variety of signals as proxies for their workplace readiness (Honea, Castro, and Peter 2017). Experience with GAI technologies during their education can signal graduates' preparedness for an AI-driven workplace. Work readiness evolves alongside technological advancements and market trends, making it important for academic curricula to develop both technical skills, such as data analytics as well as meta-skills such as creativity, critical thinking, and problem-solving (Rohm, Stefl, and Ward 2021). With the rise of generative AI, these skills are becoming even more critical, as graduates must be prepared to leverage AI tools to enhance productivity, innovation, and adaptability in an increasingly technology-driven workplace, while being mindful of the downsides of using GAI.

## EXPLORATORY REVIEW OF MARKETING SYLLABI AND FACULTY SURVEY

To better understand the role of GAI in marketing courses and examine how our proposed theoretical framework aligns with practice, we conducted a multipronged analysis, including a review of marketing course syllabi, a survey of marketing educators, and follow-up qualitative interviews. Our review of syllabi covered 11 distinct marketing courses taught by the marketing faculty at a large midwestern U.S. public university. We followed up our analysis of the syllabi with a survey of 14 full-time faculty members of the same university who teach these marketing courses. The sample of courses and instructors was chosen as a convenience sample for providing preliminary validation of our framework and should be expanded in future research (e.g., Smith, Cronley, and Barr 2012).

### Review of Marketing Syllabi and Use of AI

We first examine the learning objectives and topics extracted from the course syllabi of marketing courses offered at a large public university in the U.S. We report these for a sample of marketing courses that use AI in Table 4 for illustrative purposes.[3]

[3] The details of our sample of marketing courses as well as the objectives and topics for additional marketing courses appear in Web Appendix C.

**Table 4. Examples of Marketing Courses, Learning Objectives, and Topics Extracted from Marketing Syllabi and the Use of GAI in these Courses**

| Course | Learning Objectives | Topics | How GAI is Used |
|---|---|---|---|
| Marketing Research | Specify research questions; design a research study and select appropriate methods; collect primary data; conduct data analysis; write and present research reports. | Marketing Research Process, Research Design, Secondary Data Research, Observational Research<br>Survey Research, Measurement Scales, Questionnaire Development, Sampling Methods, Data Analysis, Research Report Writing and Presentation | Data analysis and visualization |
| New Product Development | Understand the NPD process-introduce students to the fundamental concepts, stages, and methodologies involved in developing new products, from idea generation to commercialization; guide students to identify customer needs, preferences, and pain points and derive insights to create products that address market needs; combine appropriate tools and frameworks, resulting in new product ideas; design a new product/service experience. | Product Launch, Design Thinking, Prototyping, Concept Testing | Project-based learning and ideation |
| Digital Marketing | Develop critical thinking and writing skills; learn to make strategic decisions via the process of research, analysis, thought and informed judgment; learn how to research brands, develop strategies, determine problems and solutions, evaluate information and present and defend your work; increase your level of marketing competence and professionalism by creating an environment that helps you sharpen the following skills: information gathering and analysis; inductive and deductive thinking; written and verbal communication; organization and planning; time management; interpersonal/teamwork; and presentation. | Digital & Social Media Marketing, Content Marketing, Influencer Marketing, SOSTAC Digital Marketing Model, Social Media Marketing, Website Development, Search Engine Marketing, Email Marketing, Mobile Marketing, Marketing Analytics | Content generation (e.g., syllabus, case studies), student engagement (e.g., discussion prompts), and assessment (e.g., feedback) |
| Marketing Analytics | Understand different concepts in Marketing Analytics; apply quantitative methods in R to real-world marketing scenarios; analyze, interpret and effectively communicate results from Marketing Analytics applications; create compelling presentations to showcase data and results effectively | Customer Lifetime Value Analysis, Churn Prediction, Market Basket Analysis, Conjoint Analysis, Digital & Social Media Marketing, Marketing Analytics, Statistical Analysis for Marketing, Machine Learning in Marketing, Marketing Decision Models | Data analysis (e.g., learning to code), Chatbot (e.g., course Q&A), and student engagement (e.g., asynchronous discussion forums) |

The varied learning objectives in marketing courses can shape how GAI is integrated into instruction, taking on different roles depending on the cognitive demand. For lower-order objectives, like understanding concepts, GAI can act as a *tutor*, guiding students through foundational material with explanations, quizzes, and feedback (Jürgensmeier and Skiera 2024). For higher-order tasks, where students must analyze data and develop strategies, GAI becomes a *tool*, assisting in complex tasks like data analysis or simulating market scenarios (Mollick and Mollick 2023, Rindfleisch, Kim, and Lee 2024). In collaborative projects, such as creating brand communication campaigns, GAI may serve as a *teammate*, helping students brainstorm ideas or evaluate campaign effectiveness, fostering creativity and critical thinking (Meincke et al. 2024; Hwang et al. 2020). The dynamic use of GAI in various roles within and across marketing courses supports a more personalized and effective learning experience, tailored to specific course objectives. To learn about GAI's role in this diverse set of courses, we conducted follow-up surveys and discussions with the instructors about their AI usage and mapped their responses to the type of course and course objectives.

**Survey of Marketing Educators and Results**

To understand the use of GAI in marketing courses, we surveyed the course instructors. Our survey respondents included tenured, tenure-track, and specialized faculty members from a marketing department. Note that unlike Guha, Grewal, and Atlas (2024) who focus on early perception of AI use among marketing faculty, the emphasis of our survey was to understand *how* AI is being used, in what roles, for what goals and topics, and how effective educators perceive it to be. The questions in the survey included (a) the courses taught by respondents to map them to the syllabi, (b) their familiarity with AI technologies (1 = not at all familiar, 5 = extremely familiar), (c) whether they currently incorporate AI technologies in their marketing education, (d) specific areas of marketing education where AI is utilized, (e) the effectiveness of AI technologies in enhancing student learning (rated on a scale of 1 = not at all effective, 5 = extremely effective), and (f) the perceived effectiveness of AI in various roles, such as *tutor, teammate,* and *tool.*

The responses revealed that 42% of the marketing educators consider themselves to be very or extremely familiar (i.e., ratings of four and above) with GAI technologies and that 64% of marketing educators incorporate AI in their courses. In terms of how they use *AI,* most use GAI as a *tool*, followed by *teammate* and *tutor,* depending on the course objectives. The use of GAI in these roles is also illustrated by the following response from an instructor of the Digital Marketing course:

> *"I'm excited to adopt AI as a tool to enhance my teaching and, more specifically, to generate contemporary examples that illustrate the concepts I'm teaching. I allow students to use generative AI primarily as an expert, research assistant, proofreader, peer reviewer, tutor, and cultural coach to support their learning and academic tasks."*

Our survey further revealed that instructors of higher-level courses (e.g., 400- or 500-level courses) and elective courses (e.g., Marketing Analytics) perceive GAI to be more effective when used as a *teammate*, while those of lower-level courses (e.g., 300-level courses) and core courses (e.g., Marketing Management) perceive GAI to be more effective when used as a *tutor* or *tool*. We found support for this pattern from our educator interviews through the following quotes:

> *"I created a chatbot for a marketing simulation to help answer student questions. I also allow students to use AI as they see fit throughout the course. I don't explicitly tell them to use AI as a tutor, tool, or teammate but based on how I integrate it, AI would probably best fall under the tool category in my Marketing Management core course."*

> *"In my Marketing Analytics elective course, I ask students to use AI in more of a teammate role when working on their group final projects. The project requires students to collect data, formulate their own analytics question, and report their findings. AI plays a crucial role in helping students brainstorm questions, analyze data, and even refine their reports. It's not* *just a tool for them; it's a collaborative partner in the process. For example, if they come up with three possible analytics problems, I encourage them to assess using AI which one is most feasible and interesting before they get my feedback. Ultimately, I want students to submit their own work but during the process, improve upon it using AI as a partner."*

These responses highlight how instructors teaching foundational and core courses like Marketing Management tend to view GAI as a practical tool for addressing operational challenges and streamlining course management rather than as a collaborative teammate in advanced and elective courses like Marketing Analytics. These responses also highlight that GAI's role in a course takes into account the course level and objectives (e.g., higher-level courses could encourage more dynamic and collaborative uses of the AI technology).

We further examined patterns of GAI usage and emphasis on GAI skills in courses with higher- vs. lower- order learning objectives. First, we identified higher-order learning objectives from course syllabi as those that contained words like "create" or "develop" rather than "understand" or "learn" (Bloom et al. 1956). Next, we cross-referenced course objectives and the GAI use reported by the instructors in the survey. We found that courses requiring higher-order thinking in their course objectives (e.g., objectives containing words like "create" or "develop") tend to use and emphasize GAI skills more in any of its roles. Importantly, we analyzed the instructor-provided ratings of the effectiveness of AI in these roles based on the high- vs. low-order of the marketing course. The results appear in Figure 1 and show that AI is perceived most useful as a *teammate* in higher-level courses but as a *tutor* or *tool* in lower-level courses based on the proportion of respondents who rated it a 4 or 5 on a 5-point effectiveness scale.

**Figure 1. Survey Results on the Role of AI as Tutor, Teammate, and Tool**

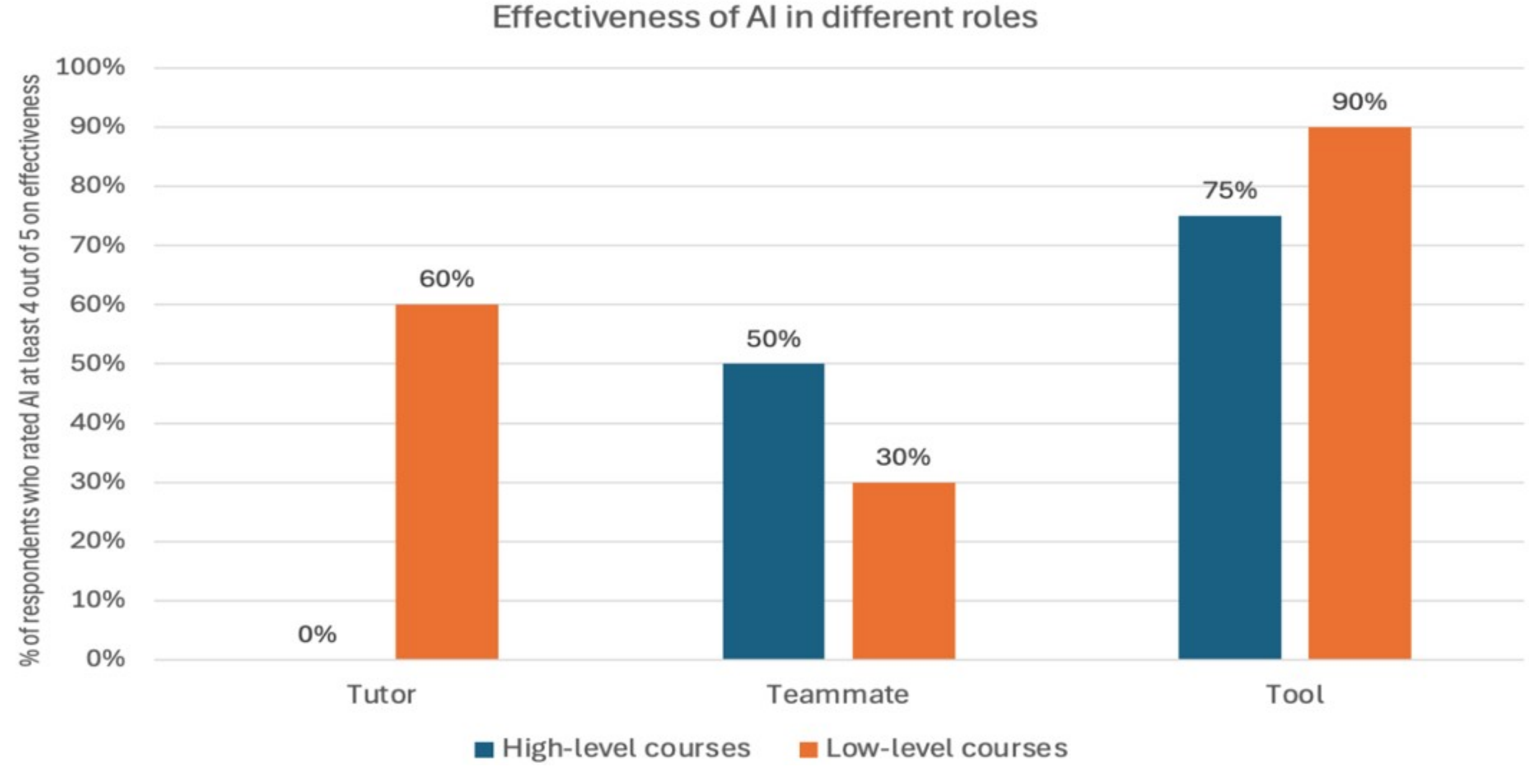

*Notes:* High-level courses refer to 400- and 500-level courses, such as *Marketing Management* and *Marketing Analytics*. Low-level courses refer to 300-level courses, such as *Principles of Marketing* and *Marketing Research*. N=14 instructors.

In addition to assessing how GAI is used in different marketing courses and its overall effectiveness, we also examined faculty perceptions of AI's effectiveness. We report these results in Table 5. This table shows that GAI was rated highly for explaining concepts and enhancing the clarity of course material, with 42.86% of respondents perceiving it as *highly effective* in facilitating teaching presence. It was also viewed as beneficial in promoting social presence by facilitating engagement and interactions among learners, with 35.71% of respondents rating it *highly effective* on this dimension. Moreover, GAI demonstrated significant effectiveness in enhancing cognitive engagement and supporting learners' critical thinking, with 85.71% of respondents rating it *highly effective* for its role in promoting cognitive presence and deepening content interactions.

**Table 5. Survey Results on the Effectiveness Ratings of GAI in Teaching, Social, and Cognitive Presence**

| Aspect | Description | Rating (%) (4 or 5) |
|---|---|---|
| Teaching Presence | GAI effective in explaining concepts and enhancing understanding | 42.86 |
| Social Presence | GAI facilitating engagement and learner interactions | 35.71 |
| Cognitive Presence | GAI supporting critical thinking and content interaction | 85.71 |

**Example Assignment: GAI Use by Marketing Educators in Various Roles**

Consider a social media marketing assignment focused on analyzing viral marketing campaigns as part of a course on Social Media and Digital Marketing. Marketing educators can modify the assignment and leverage GAI in distinct roles—as a tutor, teammate, and tool—as shown in Table 6.[4]

**Table 6. Marketing Assignment Example: Modification for GAI Roles**

| GAI Role | Description | Assignment Instruction Modifications |
|---|---|---|
| Tutor | Students independently identify viral marketing campaigns and then engage in a structured dialog with GAI as an expert guide. GAI uses a Socratic approach to ask probing questions that deepen students' understanding of theoretical concepts. | 1. Identify three viral marketing campaigns individually.<br>2. Use GAI as a tutor to evaluate each campaign.<br>3. Respond to GAI's guiding questions that focus on theoretical concepts like social transmission, emotional resonance, and KPIs. |
| Teammate | Students collaborate with GAI as a partner to identify and evaluate campaigns. GAI offers complementary perspectives, challenges assumptions, and suggests alternative viewpoints, encouraging joint ideation and analysis. | 1. Work with GAI to identify and analyze three viral marketing campaigns.<br>2. Engage in dialogue where GAI offers perspectives or challenges assumptions.<br>3. Incorporate insights from GAI into your evaluation while maintaining a critical approach. |
| Tool | Students use GAI as a resource to gather initial campaign data, analyze social media metrics, or compile preliminary insights. The assignment specifies appropriate use cases while requiring independent analysis and recommendations. | 1. Use GAI to gather data, such as social media metrics or campaign themes for three marketing campaigns.<br>2. Focus on using GAI for research and analysis tasks.<br>3. Ensure that your final analysis and recommendations are independently formulated. |

[4] Additional details about the assignment appear in Web Appendix D along with potentials ways and system prompts for training GAI in desired roles (i.e., as a *tutor, teammate*, or *tool*).

*Example of GAI as a Tutor.* In this implementation of the assignment, students first independently identify potential viral marketing campaigns. They then engage with GAI in a structured dialogue where it serves as an expert guide, helping them evaluate their selected campaigns against established theoretical frameworks. For instance, when a student proposes analyzing a campaign, the GAI tutor might ask: "How does this campaign exhibit the key characteristics of viral marketing we discussed, such as social transmission and emotional resonance?" Through this Socratic approach, GAI helps students deepen their understanding of theoretical concepts while developing critical analysis skills.

*Example of GAI as a Teammate.* This implementation of the assignment structure evolves to emphasize joint analysis and ideation. Students and GAI work together to identify and evaluate campaigns, with GAI offering complementary perspectives and challenging assumptions. For example, if a student identifies a campaign's viral elements, GAI might respond: "While you've noted the emotional appeal, let's also consider how the campaign's timing and platform choice contributed to its virality. What if we examined..." This collaborative approach helps students develop both analytical and teamwork skills while maintaining their agency and developing their own aptitude in the learning process.

*Example of GAI as a Tool.* In this implementation of the assignment, GAI serves primarily as a research and analysis resource. Students use GAI to efficiently gather initial campaign data, track social media metrics, or generate preliminary content analyses. The assignment explicitly specifies appropriate use cases, such as: "You may use GAI to compile social media engagement statistics and identify common themes across campaigns, but your analysis and recommendations must be your own." This implementation helps students learn to leverage AI for specific analytical tasks while developing independent critical thinking skills.

These three implementations demonstrate distinct pedagogical objectives. The *tutor* role seeks to develop students' theoretical understanding and analytical capabilities and enables teaching and cognitive presence. The *teammate* role seeks to improve collaborative problem-solving and critical thinking, primarily enhancing social presence. The *tool* role builds proficiency in using AI for specific analysis tasks, focusing predominantly on cognitive presence.

Through specific instructions to the students and the AI in its various roles, instructors can aim to maintain academic rigor while preparing students for different aspects of AI integration in marketing practice. This example illustrates how thoughtful assignment design can leverage GAI's capabilities while advancing specific learning objectives within our theoretical framework. While Table 6 illustrates how GAI can be integrated into a social media marketing assignment, its applications extend beyond viral marketing analysis. For example, GAI could be leveraged in a consumer behavior course to analyze sentiment trends, assisting students in evaluating emotional triggers in advertising campaigns. Similarly, in a marketing strategy course, students could collaborate with GAI to assess competitive positioning and market segmentation strategies based on real-time industry insights. These varied implementations highlight GAI's flexibility in enhancing assignments across multiple marketing domains.

Additionally, considering the broader 4Ps of marketing (Product, Price, Place, Promotion) in a *Principles of Marketing* course, GAI's roles can be extended across marketing assignments in meaningful ways. In product development, GAI can act as a *teammate*, helping students brainstorm new product ideas, generate consumer personas, and evaluate potential gaps in the market. For pricing strategies, GAI serves as a *tool*, allowing students to analyze historical pricing models, elasticity trends, and competitive pricing through data-driven insights. When studying distribution and place, GAI can function as a *tutor*, prompting students to critically assess logistics, supply chain considerations, and omnichannel strategies through guided questioning. Finally, in promotional strategy, GAI can be used as both a *teammate* and *tutor*, encouraging students to refine campaign ideas, assess past promotional successes, and optimize creative execution. Irrespective of the nature of assessment and course, for each of the GAI roles, the instructor provides instructions (i.e., system prompts) to the GAI chatbot. Web Appendix D lists the system prompts corresponding to each GAI role.

So far, we have shown that the role GAI assumes in a marketing course based on the course objectives can influence the learners' educational experience through *teaching, social,* and *cognitive* presence (Wang et al. 2024). Next, we discuss the potential consequences of GAI use depending on its three roles in our framework.

## CONSEQUENCES OF GAI USE BY MARKETING EDUCATORS

While marketing educators are integrating GAI into their courses in various ways, there is substantial uncertainty about its potential impact on learners and the workplace. In addition to excitement about these new technologies, there is a fair amount of skepticism. This is highlighted in the following response from a marketing faculty member we interviewed:

*"I'm not entirely sure about the consequences [of GAI], but I have had some concerns from observing my students use GAI. First, GAI may make students less capable in terms of innate human skills, such as independent problem-solving, potentially hindering their ability to function autonomously. While it saves them time and effort by performing tasks on their behalf, it could also diminish their motivation and interest in actively engaging with the learning process. Furthermore, many students are using AI at a rudimentary level, and it's unclear whether this will benefit them in their future careers. In fact, I worry that it could set the stage for their obsolescence, as many students were already struggling before, and the reliance on AI might only make things worse. In the end, if the optimistic view of AI leads to an age of abundance, maybe it will work out—but if not, we could face serious challenges in education and beyond."*

Marketing education is focused on developing both soft and hard skills among learners to make them workplace-ready (e.g., Freeman et al. 2017; Greenacre et al. 2017; Key, Czaplewski, and Ferguson 2019; McArthur et al. 2017). Based on our review of marketing course syllabi and survey of the course instructors, these skills can range from soft skills like teamwork to technical skills like programming and analytics. Table 7 reports the results of our survey of marketing faculty regarding the key skills their marketing courses emphasize. Most marketing courses focus on soft skills, such as communication, creativity, adaptability, problem-solving and critical thinking, storytelling, teamwork and collaboration, strategy and planning, project management, leadership, and presentation. In addition, many courses also seek to develop hard skills, such as sales prospecting and development, content creation and distribution, marketing research, digital marketing and SEO, social media marketing, marketing analytics and general technology proficiency as described in the SoTL literature on workplace readiness of marketing graduates (Kurtzke and Setkute 2021; Floyd and Gordon 1998). In applied fields like marketing, educational experience will translate to the workplace when "soft" or applied course features are enhanced, particularly creative-thinking and problem-solving skills (Arbaugh, Bangert, Cleveland-Innes 2010; DeFillippi and Milter 2009).

GAI, through its influence on the learners' educational experience in various roles as *tutor, teammate,* or *tool*, can impact learners' skills and their job-market readiness (Grewal, Guha, and Becker 2024). In this section, we discuss the potential implications of GAI when applied in its roles of *tutor, teammate*, and *tool* on important outcomes with a focus on marketing skills..

*Consequences of GAI as a Tutor.* As a tutor, GAI can transform *teaching* and *cognitive* presence by automating or augmenting instructional tasks, such as personalized feedback, adaptive learning paths, and real-time content delivery. This has clear implications for workplace readiness, as these functions align with industry needs for marketers skilled in data-driven decision-making and content optimization. For example, marketers frequently use GAI for tasks such as audience targeting, analytics, and content creation (Goldman 2024). However, over

**Table 7. Marketing Courses and Targeted Marketing Skills Identified by the Instructor**

**(a) Soft Skills**

| Course Name | Communication | Creativity | Adaptability | Problem-Solving & Critical Thinking | Story Telling | Teamwork & Collaboration | Project Management | Leadership | Presentation |
|---|---|---|---|---|---|---|---|---|---|
| **Principles of Marketing** | | X | X | X | | | | | |
| **Marketing Research** | X | X | X | X | X | X | X | X | X |
| **Marketing Communications** | X | X | X | X | X | X | X | X | X |
| **Consumer Behavior** | X | X | X | X | X | X | X | X | X |
| **Pricing Strategy** | X | X | X | X | X | X | X | X | X |
| **New Product Development** | X | X | X | X | X | X | X | X | X |
| **Brand Management** | X | X | X | X | X | X | X | X | X |
| **Digital Marketing** | X | X | X | X | | X | X | X | X |
| **International Marketing** | X | X | X | X | X | X | X | X | X |
| **Marketing Analytics** | X | X | X | X | X | X | X | X | X |
| **Advanced Marketing Management** | X | | X | X | | X | X | | X |

**(b) Hard Skills**

| Course Name | Sales Prospecting & Development | Content Creation & Distribution | Marketing Research | Digital Marketing & SEO | Social Media Marketing | Marketing Analytics & Analysis | Technology Proficiency | AI Use & Experience |
|---|---|---|---|---|---|---|---|---|
| **Principles of Marketing** | | X | X | X | X | | | |
| **Marketing Research** | | | X | | | X | | X |
| **Marketing Communications** | | X | X | X | X | | | |
| **Consumer Behavior** | | X | X | | X | | X | |
| **Pricing Strategy** | X | X | X | X | | X | | |
| **New Product Development** | | X | X | | | | | X |
| **Brand Management** | | | X | | | | | |
| **Digital Marketing** | | X | X | X | X | X | X | X |
| **International Marketing** | | | X | | | X | | |
| **Marketing Analytics** | | | X | | X | X | | X |
| **Advanced Marketing Management** | | | X | | | X | | |

*Notes:* The marketing instructors for each of the ten courses filled out the skill matrix i.e., identified which skills are used in their course.

reliance on GAI for these tasks may inadvertently deprive students of the opportunity to develop foundational skills in strategic problem-solving and critical thinking—abilities crucial for managerial roles (Horvath 2024). The example assignment in Table 6 uses GAI as a *tutor* with goals that are targeted at teaching theoretical concepts to students while learning to think deeply.

*Consequences of GAI as a Teammate.* When used as a *teammate*, GAI mainly supports *social* presence by facilitating collaboration and enhancing peer-to-peer interactions in marketing courses. This role is particularly relevant for developing teamwork, critical discourse, and problem-solving skills. For example, GAI can act as a brainstorming partner during group projects, offering suggestions for creative marketing campaigns or analyzing the competitive landscape. These activities mimic real-world marketing scenarios, where collaboration with AI tools is increasingly common (Davenport and Mittal 2022). However, ethical concerns arise, such as the potential for over-reliance on GAI to generate ideas or for AI to introduce biases into group work. Instructors should be mindful and observe these group dynamics, potentially by observing the group's chat and interactions with the GAI. The example assignment in Table 6 that uses GAI as a t*eammate*, facilitates collaborative brainstorming and asks students to refine their campaign strategies through iterative discussions rather than simply submitting GAI's outputs.

*Consequences of GAI as a Tool.* GAI as a *tool* mainly influences *teaching* and *cognitive* presence by enabling learners to execute tasks and apply new concepts to real-world problems. This role aligns closely with industry demands for technical proficiency in areas such as digital marketing, data analysis, and SEO optimization. For instance, in marketing courses, GAI tools can help students analyze market trends or develop data-driven strategies, mirroring tasks marketers perform in the workplace (Jacobs and Muro 2023). However, the success of GAI in this role depends on the educator's ability to integrate it into the curriculum effectively. The example assignment in Table 6, which uses GAI as a *tool*, assists students in analyzing data and reports, allowing them to instead focus on crafting innovative solutions and making strategic decisions.

Beyond the roles assumed by GAI, the incorporation of GAI in marketing education is posited to influence learners' perceived and actual learning outcomes, which, in turn, affect their work readiness (Grewal, Guha, and Becker 2024). While marketing educators strive to nurture higher-order skills, such as critical thinking, creativity, and strategic problem-solving, an over-reliance on GAI might hinder the development of these competencies (Horvath 2024). Students who depend heavily on GAI for tasks like content creation or data analysis may neglect opportunities for deep learning and critical evaluation. This could have long-term consequences for their career success. However, when GAI is leveraged appropriately, it can enhance students' learning experiences and skill acquisition. GAI can facilitate improved engagement and productivity, providing learners with a competitive advantage in the job market. The familiarity with GAI technologies equips graduates with the necessary skills. In Web Appendix E, we provide additional concrete examples of GAI use from a marketing course to provide educators a template.

While the benefits can be substantial, integrating GAI into marketing education presents several challenges. Next, we discuss key ethical considerations for the use of GAI.

## ETHICAL CONSIDERATIONS IN GAI INTEGRATION

The integration of GAI in marketing education raises important ethical considerations that educators and institutions must address (Ferrell and Ferrell 2024, Hermann and Puntoni 2024). These considerations encompass privacy, equity, academic integrity, transparency, and potential overreliance on AI systems,  each requiring careful attention to ensure responsible implementation of GAI across its roles as a *tutor, teammate,* and *tool.*

Privacy and data protection concerns are paramount when implementing GAI in education. Student interactions with GAI systems, particularly in its role as a *tutor* or *teammate,* generate substantial data about learning patterns, comprehension levels, and personal perspectives. This data collection raises questions about compliance with educational privacy regulations like FERPA, especially regarding data storage, usage rights, and potential sharing with third-party GAI providers. Marketing educators must establish clear protocols for data handling and ensure that GAI implementations maintain student privacy while supporting learning objectives. If the institution has invested in infrastructure that provides privacy-preserving access to students and faculty, it will be easier for faculty to use GAI tools. If instructors need to make their own arrangements for access, they will need to provide clear guidelines to students on how to use public GAI technologies in a privacy-preserving manner.

Equity and access considerations are equally crucial. The digital divide in access to GAI tools can create disparities in educational opportunities. Some students may face barriers due to costs associated with premium GAI services or limited access to required technology infrastructure. Moreover, existing biases in GAI systems may disadvantage certain student groups, particularly in language processing and feedback generation. Marketing educators must develop inclusive implementation strategies that ensure all students can benefit from GAI integration, regardless of their technological or economic circumstances. Several providers offer free access to GAI tools, along with open-source models that provide capabilities comparable to leading AI technologies.

Academic integrity in GAI-enhanced marketing education requires careful balance. While GAI can enhance learning through personalized tutoring, collaborative ideation, and analytical support, it also presents challenges in maintaining the authenticity of student work. Clear guidelines needed to distinguish between appropriate GAI assistance and academic misconduct. For instance, using GAI as a *tool* for data analysis in marketing research projects may be appropriate, while using it to generate complete campaign strategies independently may not. Educators must develop and communicate clear policies about acceptable GAI use across its various roles. Here is an excerpt from AI use guidelines used by one of the instructors we surveyed which has been vetted by the teaching and learning unit in the college.

*"Unless otherwise indicated, you may use generative AI (e.g., ChatGPT, Copilot, Gemini, Claude, or other chatbots) in the following roles: Sounding board, Expert, Draft Writer/outline generator, Language Translator, Conversation Partner, Devil's Advocate, Research Assistant, Proofreader, Peer reviewer, Tutor and Cultural Coach. Note, however, that usage of generative AI outside of these parameters could result in a violation of the university student code and potentially, dismissal from the university. I highly encourage you to speak freely with me about your usage of AI; it's not a taboo topic for me. My goal is for us to freely and openly discuss generative AI and its ethical usage early and often throughout the semester."*

Transparency and trust building are essential for successful GAI integration. Marketing educators should clearly communicate how GAI is used in their courses, including its limitations and potential biases (Puntoni 2024). Students should understand when they are interacting with GAI systems, whether as a *tutor* providing feedback, a *teammate* in collaborative projects, or a *tool* for analysis. This transparency helps build trust in GAI-enhanced education while ensuring students maintain realistic expectations about GAI capabilities.

Beyond privacy, equity, and academic integrity concerns, marketing educators must also consider the risk of overreliance on GAI creating a 'paradox of automation' where automation of routine tasks may prevent students from developing fundamental skills needed for more complex tasks (Madhavaram and Appan, 2025). For example, if basic aspects of marketing analysis become automated through GAI, students may not develop the deep understanding and critical thinking abilities required for strategic marketing decisions. This can become particularly problematic if GAI systems fail or produce unreliable outputs, leaving students without the core competencies to independently evaluate and solve marketing problems. Marketing educators need to carefully balance GAI integration with opportunities for students to develop and practice essential marketing skills independently. In addition to ethical concerns, adoption barriers to the use of GAI must also be considered. Future research can examine these open issues. Next, we propose a future research agenda based on our framework.

## RESEARCH AGENDA AND FUTURE DIRECTIONS

Our study is among the first to provide a systematic review and integrative perspective on emerging issues related to GAI in marketing education. This section outlines several directions for future research, summarized in Table 8, alongside illustrative questions. In addition to identifying key research avenues within the scope of our research, we also provide examples of questions that future research can address, such as understanding of GAI's roles in education while addressing adoption barriers and ethical considerations inherent in its implementation.

First, our research examines AI's role as a *tutor*, which can predominantly influence teaching and cognitive presence in marketing courses (see Table 3). While there is an emerging body of literature on the use of GAI in marketing education (e.g., Ding, Dong, and Grewal 2024; Grewal, Guha, and Becker 2024), the specific implications of AI's role as a *tutor* within marketing education remain largely unexplored. Future research can address how GAI tutoring influences marketing educators' ability to design personalized and adaptive marketing courses with a focus on helping learners understand and grasp key marketing concepts. Future research can also empirically examine data across courses, if available, to understand how GAI's role as a tutor varies for higher- and lower-level courses and elective- and core-courses. Additionally, questions remain about the extent to which GAI tutoring fosters or undermines trust and personal connections among learners. Another important avenue for research is the impact of GAI on asynchronous versus synchronous learning. Could the integration of GAI help bridge the engagement gap in asynchronous courses by providing real-time, adaptive feedback, making these courses feel more interactive and comparable to live instruction? What are the implications of such advancements for the effectiveness and popularity of asynchronous marketing education? Additionally, how does GAI tutoring support or hinder the four phases of cognitive presence (triggering event, exploration, integration, and resolution) through underlying process mechanisms and overall learner outcomes in marketing education? Ethical considerations, such as ensuring transparency in AI feedback and maintaining equitable access for diverse learners, also merit further investigation. Adoption barriers, including educators' apprehension about AI replacing traditional teaching roles and the lack of professional training for AI-supported pedagogy, require targeted exploration. These issues represent important future directions to understand GAI's role as a tutor in marketing education.

Second, our research explores AI's function as a *teammate*, focusing on its potential to facilitate collaboration and enhance social presence by enabling peer-to-peer interactions in marketing courses. Although early research touches on GAI's role in online learning communities (e.g., Narang, Ahlbom, and Banerjee 2024), its specific impact on group dynamics and critical discourse in marketing education is insufficiently understood. Future studies can explore how GAI as a *teammate* shapes the educator's design of collaborative activities and affects group cohesion and communication among marketing students. What are the effects of GAI teammates on the formation of group cohesion and open communication in marketing learning communities? Importantly, how can group projects be effectively assessed when GAI acts as a brainstorming partner or teammate? What new metrics or assessment methods might be needed to evaluate individual learning and skill development in the GAI era? How can instructors fairly allocate credit to learners while accounting for GAI's contributions? Ethical concerns, such as ensuring GAI promotes genuine collaboration rather than dependency, and the potential for AI-generated bias in group settings, require further scrutiny. Barriers to adoption, including limited familiarity with integrating AI into teamwork and the absence of robust ethical guidelines for its use, also need attention. These are ripe areas for future investigation.

Third, our research identifies GAI's role as a *tool,* which can influence all three presences —*teaching, social,* and *cognitive*—through its advanced analytical and content-generation capabilities. While the literature on technology-enhanced learning is extensive (e.g., Mayer 2019), GAI's specific role in transforming course content and delivery in marketing education remains underexplored. Future research can examine how GAI tools support educators in structuring learning experiences and tailoring materials to diverse student needs. Furthermore, questions arise about how educators can train students to critically evaluate AI outputs and use them effectively to solve marketing problems. Ethical considerations, such as safeguarding data privacy and preventing misuse of AI-generated content, are central to this role. Barriers to adoption, including lack of access to advanced AI tools, insufficient institutional support, and concerns about intellectual property and plagiarism, must also be addressed. Developing guidelines and course policies for specific ways in which AI use is permissible also requires an understanding of the impact of these tools on learning.

Beyond examining current implementation challenges, future research must also anticipate and prepare for emerging developments in AI technology that could reshape GAI's roles in marketing education. Advances in natural language processing may enhance GAI's ability to provide more nuanced, contextually aware feedback as a tutor, potentially approaching human-like capabilities in understanding student learning patterns and adapting pedagogical approaches accordingly. Improvements in machine learning algorithms could enable more sophisticated collaborative capabilities when GAI acts as a *teammate*, particularly in creative tasks like campaign development or market analysis. Additionally, emerging multimodal AI systems that integrate text, voice, and visual processing could transform GAI's effectiveness as a tool, enabling more comprehensive analysis of marketing materials and consumer behavior data.

These technological developments raise important additional research questions about the evolution of GAI's role in marketing education. How might advances in natural language understanding affect GAI's ability to foster meaningful social presence in marketing courses, particularly in facilitating more natural peer-like interactions? What new opportunities and challenges might emerge as GAI systems become more capable of handling complex, contextualized marketing problems that currently require significant human expertise? How should marketing educators prepare for and adapt to potential paradigm shifts in GAI capabilities, particularly in maintaining the balance between technological enhancement and core marketing competency development? What implications might improved GAI capabilities have for assessment methods, particularly in evaluating students' mastery of marketing concepts versus their ability to effectively leverage AI tools? Understanding the trajectory of these technological developments is crucial for developing forward-looking frameworks that can adapt to evolving GAI capabilities while maintaining pedagogical effectiveness in marketing.

Our framework assumes the adoption of AI by educators and learners and is somewhat agnostic to the ethical barriers associated with its use. However, there may be barriers to AI adoption that future research can also further investigate. Specifically, more research is needed to explore the factors that influence marketing educators' decisions to adopt GAI, including individual, institutional, and technological drivers. These antecedents could include educators' familiarity with AI tools, perceived ease of use, and institutional support for integrating AI into curricula. The barriers to adoption are equally critical and may include concerns about the accuracy and reliability of GAI (Acar 2024), lack of training or clear guidelines for implementation, and ethical concerns such as bias, plagiarism, and data privacy. Additionally, the obstacles to adoption may vary depending on GAI's role—whether it is being used as a *tutor, teammate,* or *tool*. For example, using GAI as a *tutor* might face resistance due to concerns about replacing traditional instructional roles, while its role as a *tool* may be hindered by a lack of technical infrastructure or resources. Since many marketing educators currently rely on trial-and-error approaches to implement GAI, future research could also examine how structured frameworks or professional development programs might mitigate these barriers and promote effective adoption. Identifying role-specific obstacles will help educators tailor strategies to integrate GAI more effectively into their teaching practices while addressing key challenges.

Table 8. Future Research Agenda: Marketing Educators' Use of GAI in Various Roles

| GAI Role | Illustrative Research Questions |
|---|---|
| Tutor | How does the use of GAI as a marketing tutor influence the educational experience of marketing learners in terms of the teaching, social, and cognitive presence? How does it impact their learning and work readiness?<br>In what ways does GAI as a tutor affect the educator's role in guiding discourse and providing direct instruction in marketing education and its impact on learner outcomes?<br>How does the introduction of GAI as a tutor impact students' ability to project their personal characteristics and present themselves authentically?<br>What effect does GAI as a tutor have on trust and connection among learners in marketing courses?<br>Can GAI as a tutor make asynchronous courses more interactive, creating a learning experience closer to synchronous courses? If so, how?<br>How does GAI as a tutor support or hinder the four phases of cognitive presence (triggering event, exploration, integration, and resolution) in marketing education?<br>What role do adoption barriers—such as concerns about replacing human instructors, lack of training, and institutional resistance—play in the integration of GAI as a tutor? |
| Teammate | How does the use of GAI as a marketing teammate influence the educational experience of marketing learners in terms of teaching, social, and cognitive presence? How does it impact their learning and work readiness?<br>In what ways does GAI as a teammate affect educators' design and facilitation of group projects and collaborative learning activities?<br>How does interaction with GAI as a teammate affect group cohesion, open communication, and the exploration of diverse marketing perspectives?<br>How does collaboration with GAI as a teammate impact students' expression of their personal identities and perspectives in marketing team projects? What are the challenges and opportunities for student assessment? What new metrics or assessment methods might be needed to evaluate individual learning and skills?<br>How does the inclusion of GAI as a teammate in group work influence the collective exploration of marketing ideas?<br>How can GAI as a teammate promote critical thinking and analysis of marketing problems without over-reliance or reduced peer engagement?<br>How does the use of GAI as a teammate affect a marketing educator's ability to structure the overall learning process?<br>How do adoption barriers, such as lack of familiarity with AI-supported collaboration and ethical concerns about fairness and dependency, influence the use of GAI as a teammate? |
| Tool | How does the use of GAI as a marketing tool influence the educational experience of marketing learners in terms of the teaching, social, and cognitive presence? How does it impact their learning and work readiness?<br>In what ways does GAI as a tool influence the educator's capacity to provide timely and constructive feedback in marketing education? How does it impact assessment of individual student outcomes?<br>What role does GAI as a tool play in the instructor's ability to tailor course materials to diverse student needs in marketing education? What kind of guidance, training, and support can the instructor provide on using AI as a tool?<br>To what extent do GAI tools support students in connecting ideas and applying new concepts to marketing problems?<br>How do ethical concerns, such as plagiarism, data privacy, and intellectual property, shape the adoption and use of GAI as a tool in marketing education?<br>To what extent do adoption barriers, such as limited technical infrastructure and lack of institutional support, hinder the effective integration of GAI tools into marketing curricula? |

## CONCLUSION AND LIMITATIONS

The integration of GAI in marketing education offers transformative opportunities as well as challenges to impact student learning and preparedness for future marketing careers. This research advances our understanding of GAI in marketing education through a novel integration of Role Theory and the Community of Inquiry framework. Our investigation, combining syllabi analysis, faculty surveys, and qualitative interviews, shows that GAI can assume the roles of a *tutor, teammate,* and *tool* in marketing courses. Through these roles, GAI can influence the *teaching, social,* and *cognitive* presence, that together shape learners' educational experience in marketing courses and their marketing skills and competencies. In doing so, we provide concrete examples and actionable guidance for marketing educators. While GAI holds significant potential to improve engagement and provide personalized feedback, it also presents challenges that require careful implementation. By applying rich theoretical frameworks, we offer nuanced insights to help educators strategically incorporate GAI into curricula, creating an effective and balanced educational experience. Policymakers in higher education should prioritize training and ethical guidelines to ensure GAI supports both learning outcomes and broader workforce preparedness.

Our insights carry important implications for marketing education policy and practice. Marketing educators must develop comprehensive frameworks for GAI implementation that account for course-specific objectives while maintaining academic rigor and ethical standards. Furthermore, institutions should invest in faculty development programs that enhance GAI literacy and support effective usage across different roles and contexts. Our research suggests that policy frameworks should address three critical dimensions: ethical implementation guidelines, faculty development support, and assessment criteria that balance technological innovation with core marketing competencies.

Several limitations warrant careful consideration when interpreting our findings. First, our sample of marketing syllabi and faculty surveys (n=14) from a single large midwestern U.S. public university may limit the generalizability of findings to other institutional contexts, particularly those with different resources, student populations, or cultural settings. Second, the cross-sectional nature of our data collection, conducted during a period of rapid GAI evolution, may not fully capture the dynamic nature of GAI adoption in marketing education. Third, our faculty survey responses primarily reflect early adopters of GAI technology, potentially overestimating its perceived effectiveness across the broader marketing education community. Future research could address these limitations through larger-scale, longitudinal studies across diverse institutional settings, including private universities, liberal arts colleges, and international institutions. Additionally, investigating student perspectives and learning outcomes would provide a more comprehensive understanding of GAI's impact on marketing education. Future research opportunities emerge from our findings, particularly in examining longitudinal impacts of GAI integration on student learning outcomes and workplace readiness. Future research should also investigate changing GAI roles with evolving ethical, social, and policy landscapes. Additionally, scholars should investigate how GAI's roles might evolve as the technology advances and how marketing educators can adapt their pedagogical approaches accordingly. Given the rapid pace of technological change, continuous evaluation and refinement of GAI implementation strategies will be crucial for maintaining educational effectiveness and relevance in marketing education.